\documentclass[prb,twocolumn,groupedaddress,showpacs]{revtex4}

\usepackage{graphicx}

\newcommand{\CC}{{\cal C}}
\newcommand{\CH}{{\cal H}}
\newcommand{\Teff}{T_{\rm eff}}
\newcommand{\Ceff}{\CC_{3,\rm eff}}

\begin{document}

\title{Feedback and Rate Asymmetry of the Josephson Junction Noise Detector}
\author{D.~F.~Urban and Hermann Grabert}

\affiliation{Physikalisches Institut and Freiburg Institute for
Advanced Studies, Albert-Ludwigs-Universit\"at, 79104 Freiburg,
Germany}

\date{\today}

\begin{abstract}
The Josephson junction noise detector measures the skewness of
non-Gaussian noise via the asymmetry of the rate of escape from
the zero-voltage state upon reversal of the bias current. The
feedback of this detector on the noise generating device is
investigated in detail. Concise predictions are made for a second
Josephson junction as noise generating device. The strong
nonlinearity of this component implies particularly strong
feedback effects, including a change of sign of the rate asymmetry
as the applied voltage approaches twice the superconducting gap.
\end{abstract}

\pacs{85.25.Cp, 72.70.+m, 73.23.-b}

\maketitle

In the last decade there have been extensive theoretical studies
of non-Gaussian noise generated by nonlinear electronic
nanostructures\cite{Levitov,reviews}. This new field of full
counting statistics (FCS) has put forward numerous predictions
indicating that the higher order noise cumulants contain valuable
information about the electronic transport mechanisms within the
noise generating element that are not accessible from more
standard measurements of the conductance and the noise power
alone. A few years ago, first experimental observations of
non-Gaussian current noise have been
reported\cite{Reulet,Reznikov}, and more recently quantum point
contacts\cite{Fujisawa,Ensslin} and Josephson
junctions\cite{Pekola,Pothier} have emerged as promising on-chip
noise detectors.

The current experimental efforts mainly focus on reliable
measurements of the third noise cumulant ${\cal C}_3$, the
skewness of the noise. A Josephson junction (JJ) in the
zero-voltage state can detect the skewness of a noise current
passing the detector since ${\cal C}_3$ is odd under time reversal
and thus leads to an asymmetry of the rate of escape from the
zero-voltage state of the JJ when the direction of the bias
current is reversed.\cite{Pekola,Pothier} After some primary
suggestions for JJ noise
detectors\cite{Tobiska,Pekola1,Heikkila,Ankerhold1,Brosco},
concrete theoretical studies for the setup used in experiments
have started recently.\cite{Ankerhold2,Sukhorukov,Grabert}

Now, that the theoretical tools for predictions for JJ noise
detectors with realistic parameters are available, we shall
address here the feedback of the noise detector on the noise
generating device and show that this feedback can modify the rate
asymmetry, the quantity determined experimentally, quite
considerably. A detailed understanding of this feedback is
essential for reliable data analysis. But this issue is also of
relevance for all on-chip detectors, since this very concept
implies two mesoscopic devices, detector and measured device, that
interact on the same chip and thus have to be treated on the same
footing. To be concrete, we shall specifically investigate the
detection of non-Gaussian noise generated by a another JJ biased
by a voltage $V_N$ with $eV_N$ close to twice the gap $\Delta$ of
the superconductor. Because of the strong nonlinearity of the
noise generating device in this region, feedback of the detector
leads to particularly pronounced modifications of the rate
asymmetry.

\begin{figure}
    \begin{center}
        \includegraphics[scale=0.4]{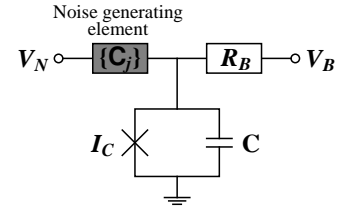}
        \caption{\label{fig:circuit} Circuit diagram of a JJ noise detector:
        A JJ with critical current $I_c$ and capacitance $C$ is biased in a
        twofold way. The branch to the right puts an Ohmic resistor $R_B$
        is series with the junction and is biased by the voltage $V_B$.
        The branch to the left is biased by a voltage $V_N$ and includes
        the noise generating nonlinear element characterized by its
        cumulants
        $\CC_j$.}
\end{center}
\end{figure}

A typical realization of a JJ noise detector is sketched in
Fig.~\ref{fig:circuit}: A JJ with capacitance $C$ and critical
current $I_c$ is connected by two branches which act as noise
sources. A bias voltage $V_B$ is applied to the first branch with
an Ohmic resistor $R_B$ in series with the junction. This branch
allows to control the detector by fixing the average bias
current, applying current pulses, and reading out voltage
signals. A second voltage $V_N$ is applied to another branch
which includes a non-linear noise generating conductor, again in
series with the detector. This noise generating element can most
generally be characterized by its charge transfer statistics,
respectively its voltage dependent current cumulants
$\{\CC_1(V),\CC_2(V),...\}$. Current experimental set-ups are
more sophisticated, but the circuit diagram in
Fig.~\ref{fig:circuit} captures the essentials of a JJ on-chip
noise detector.

The state variables of the JJ detector are the charge $Q$ on the
junction capacitance $C$ and the phase difference $\varphi$
between the order parameters of the superconductors on either side
of the tunnel barrier. The phase dynamics of the detector can be
described by classical physics, since the effective temperature
determined by the strength of Gaussian noise from the two branches
(see below) is typically much larger that the crossover
temperature below which macroscopic quantum tunneling of the
phase\cite{CL} becomes relevant. The stochastic dynamics of the
setup can then be described in terms of a Hamiltonian ${\cal
H}$\cite{GGG} which depends on the state variables $Q,\varphi$ and
the conjugate thermodynamic forces $\lambda,\mu$. We note that the
Hamiltonian of the stochastic theory has to be distinguished from
the Hamiltonian of the microscopic theory. $\CH$ depends only on
gross variables of the system and is measured in the same units as
the rate of change of entropy [$k_B\omega$] rather than energy
[$\hbar\omega$]. For the setup in Fig.~\ref{fig:circuit}
$\CH=\CH_D+\CH_B$, where
\begin{equation}
    \CH_D= \frac{2e}{\hbar}\frac{Q}{C}\mu-I_c\sin(\varphi)\lambda
    \label{HD}
\end{equation}
describes plasma oscillations of the JJ detector, and
\begin{eqnarray}\label{HB}
    \CH_B&=&\left[\frac{1}{R_B}\left(V_B- \frac{Q}{C}\right)+\CC_1\right]\lambda\\
    &+&\left[\frac{T}{2R_B} + \frac{\CC_{2}}{4k_B}\right]\lambda^2
      +2k_B\sum_{j=3}^{\infty}\frac{\CC_{j}}{j!}\left(\frac{\lambda}{2k_B}\right)^j
      \nonumber
\end{eqnarray}
adds the two biasing branches. While the branch with Ohmic
resistor $R_B$ generates Johnson-Nyquist noise and has only two
nonvanishing cumulants $\CC_{1}^B=\left(V_B- Q/C\right)/R_B$ and
$\CC_{2}^B=2k_BT/R_B$, the other branch is characterized by the
cumulants $\CC_j$ that are taken at voltage $V_N- Q/C$.

The canonical equations of motion following from the Hamiltonian
$\CH$ determine the most probable path between a given initial
state $i$ and a final state $f$, and the transition probability
between these states may be written as a path
integral\cite{GGG}
\begin{equation}\label{pi}
   p_{i\rightarrow f}
   =\int D[Q,\varphi,\lambda,\mu] \exp{\left\{-\frac{1}{2k_B} A[Q,\varphi,\lambda,\mu]\right\}}\, ,
\end{equation}
where the action functional is given by
\begin{equation}\label{action}
   A[Q,\varphi,\lambda,\mu]=\int_0^t ds
    \left[\dot{Q}{\lambda}+\dot{\varphi}{\mu}- {\cal H}(Q,\varphi,\lambda,\mu)\right]\, .
\end{equation}

During escape of the detector from the zero-voltage state, the
dimensionless quantity $e\lambda/k_B$ can be shown\cite{Grabert}
to be always much smaller than 1. Furthermore, the voltage across
the JJ detector
\begin{equation}\label{vj}
    V_J=\frac{Q}{C}=\frac{\hbar}{2e}\dot\varphi
\end{equation}
that builds up under noise activation is proportional to
$\lambda$. Keeping in Eq.~(\ref{HB}) only terms that are at most
of third order in $\lambda$ and $Q/C$, we obtain ${\cal H}=
{\CH}_2+{\CH}_3$, where
\begin{equation}
\label{eq:H2}
    {\CH}_{2}=\frac{2e}{\hbar}\frac{Q}{C}\mu  -I_c\sin(\varphi)\lambda +\left(I_{t}- \frac{1}{R_{\vert\vert}}\frac{Q}{C}\right) \lambda
 +\frac{\Teff}{2R_{\vert\vert}}\lambda^2
\end{equation}
describes the stochastic dynamics in presence of Gaussian noise,
while
\begin{eqnarray}
\label{eq:H3}
   {\CH}_3&=&\frac{1}{24k_B^2}\CC_{3}^N \lambda^3
   - \frac{1}{4k_B}\frac{\partial \CC_{2}^N}{\partial V_N}\frac{Q}{C}\lambda^2
   + \frac{1}{2}\frac{\partial^2 \CC_{1}^N}{\partial
   V_N^2}\frac{Q^2}{C^2}\lambda\quad
\end{eqnarray}
includes the leading order effects of non-Gaussian noise. Here we
have defined the total bias current through the detector
\begin{equation}
    I_{t}=\frac{V_B}{R_B}+\CC_{1}^N
\end{equation}
and the parallel (differential) resistance
\begin{equation}\label{parallelr}
  \frac{1}{R_{\vert\vert}}=\frac{1}{R_{B}} + \frac{\partial \CC_{1}^N}{\partial V}
\end{equation}
of the circuit. Moreover, we have introduced the effective
temperature
\begin{eqnarray}
\label{eq:Teff}
   \Teff&=&R_{\vert\vert}\left[\frac{T}{R_B} + \frac{\CC_{2}^N}{2k_B} \right]
\end{eqnarray}
characterizing the strength of Gaussian noise. The $\CC_j^N$ are
the cumulants $\CC_j$ taken at voltage $V_N$.

In terms of the effective parameters introduced above, the leading
order Hamiltonian ${\CH}_2$ has the standard form for a biased JJ
described by the resistively and capacitively shunted junction
model.\cite{RCSJ} The phase $\varphi$ moves in the ``tilted
washboard" potential
\begin{equation}\label{potential}
   U(\varphi)=-\frac{\hbar I_c}{2e}\left[\cos(\varphi)+s\varphi
   \right]
\end{equation}
driven by Gaussian noise. Here $s=I_t/I_c$ is the dimensionless
bias current. For $0<s<1$, the potential has extrema in the phase
interval $[0,2\pi]$ at
\begin{equation}\label{well}
   \varphi_{\rm well, top}=\arcsin(s)=\frac{\pi}{2}\mp \delta\, ,
\end{equation}
with $\delta\approx\sqrt{2(1-s)}$  for $1-s\ll 1$. When the JJ is
trapped in the state $\varphi_{\rm well}=\frac{\pi}{2}-\delta$,
the average voltage $V_J$ across the junction vanishes. This
zero-voltage state is metastable, and to escape from the well,
the junction needs to be activated to the barrier top at
$\varphi_{\rm top}=\frac{\pi}{2}+ \delta$ by noise forces. The
weak non-Gaussian noise described by ${\CH}_3$ also gives a
contribution to this process, and the rate of escape $\Gamma$ may
be written as
\begin{equation}\label{rate}
   \Gamma = f\,{\rm e}^{-\left(B_2+B_3\right)}\, ,
\end{equation}
where $f$ is the prefactor of the rate, and the exponential factor
is determined by the action of the most probable escape path,
which is a solution of the canonical equations of motion following
from the Hamiltonian with the boundary conditions
$\varphi(t=-\infty)=\varphi_{\rm well}$ and
$\varphi(t=+\infty)=\varphi_{\rm top}$. The exponential rate
factor has a dominant contribution
\begin{equation}\label{b2}
   B_2=\frac{\Delta U}{k_B\Teff}\,
\end{equation}
arising from ${\cal H}_2$, where
\begin{equation}\label{DU}
    \Delta U = U(\varphi_{\rm top}) -U(\varphi_{\rm well})
\end{equation}
is the barrier height of the metastable well. The correction $B_3$
due to non-Gaussian noise may be evaluated by treating $\CH_3$ as
a perturbation. Following the lines of reasoning of Ref.\
\onlinecite{Grabert} one finds
\begin{equation}\label{b3}
   B_3=
-\frac{1}{\left(k_B\Teff\right)^3} \left(\frac{\hbar}{2e}\right)^3
\Ceff J \, ,
\end{equation}
where
\begin{equation}
\label{eq:C3eff}
  \Ceff=\CC_{3}^N- 3k_B\Teff\frac{\partial \CC_{2}^N}{\partial V_N}
  + 3(k_B\Teff)^2\frac{\partial^2 \CC_{1}^N}{\partial V_N^2}
\end{equation}
is the \emph{effective} third noise cumulant. The second and
third terms in Eq.~(\ref{eq:C3eff}) arise from the feedback of
the JJ on the noise generating device, which is a consequence of
the finite voltage $V_J$ arising during escape. The quantity
\begin{equation}\label{jrelax}
   J=-\frac{1}{6}\int_{-\infty}^{\infty} dt \, \dot\varphi_{\rm
   relax}^3(t)
\end{equation}
is expressed in terms of the deterministic trajectory
$\varphi_{\rm relax}(t)$ starting at $t=-\infty$ at the barrier
top $\varphi_{\rm top}$ and relaxing (in the absence of noise
forces) to the well bottom $\varphi_{\rm well}$ reached at time
$t=\infty$.

The coefficient $B_2$ is even under time reversal, while $B_3$ is
odd, like the third noise cumulant. Accordingly, when the
direction of the bias current is reversed, the rate coefficient
shows an asymmetry which is a signature of non-Gaussian noise. The
deviation of the rate ratio\begin{equation}\label{asy}
    \frac{\Gamma(I_t)}{\Gamma(-I_t)} = \exp\left[-2B_3(I_t)\right]
    \approx 1 - 2B_3(I_t)\, ,
\end{equation}
from 1 allows for an experimental determination of $B_3$ despite
the fact that the corrections due to non-Gaussian noise are
typically small.

In the following we will focus on a specific noise generating
element, namely a JJ with normal state tunneling resistance $R_t$
and capacitance $C_N$ biased by a voltage $V_N$ which is tuned
close to twice the gap $\Delta$ of the superconductor. In view of
the lead to the voltage source, this noise generating element is
in series with an approximately Ohmic lead resistance $R_N$ in the
range of a few $100~\Omega$. A corresponding setup is currently
studied experimentally,\cite{PothierComm} and is well suited to
examine the feedback of the detector on the noise generating
element, since the behavior of the JJ is highly nonlinear for
$eV_N\approx 2\Delta$.

A JJ with parameters around those indicated in Tab.\
\ref{tab:Parameters} has a very low sub-gap current, and for
$eV>2\Delta$ the current is carried by tunneling quasiparticles.
We can then employ the semiconductor model,\cite{Tinkham} where
the superconductor is described in terms of quasiparticles with a
density of states (DOS) $N_S(E)$ given by the familiar BCS
expression
\begin{eqnarray}
   \frac{N_S(E)}{N_0}&=&\frac{|E|}{\sqrt{E^2-\Delta^2}}\, ,
\end{eqnarray}
where $N_0$ is the normal state DOS at the Fermi level. The mean
current through the JJ can be written as
\begin{eqnarray}
\langle I \rangle &=&\CC_1^{\rm qp}\;=\;
e(\Gamma_{\rightarrow}-\Gamma_{\leftarrow})\, ,
\end{eqnarray}
where the forward and backward quasiparticle tunneling rates
across the junction interface are given by\cite{Falci,Ingold}
\begin{eqnarray}
\label{eq:tunnelrate}
   \Gamma_{\rightarrow}(V)&=&\frac{1}{e^2R_t}\int dE dE'
   \frac{N_S(E)N_S(E'+eV)}{N_0^2}
\nonumber\\&&
   \times f(E)\left[1-f(E'+eV)\right]P(E-E')
\end{eqnarray}
and $\Gamma_{\leftarrow}(V)=\Gamma_{\rightarrow}(-V)$,
respectively. Here, $f(E)$ is the Fermi distribution. The
influence of the electromagnetic environment is described in terms
of the probability $P(E)$ that a tunneling quasiparticle looses
the energy $E$ to the environment. This dynamical Coulomb blockade
(DCB) effect leads to a smearing of the expected discontinuity of
$I(V)$ at $eV=2\Delta$. Since the phase dynamics of the JJ
detector is very slow on the time scale of quasiparticle tunneling
in the noise generating JJ, the impedance of the electromagnetic
environment is essentially given by the resistance $R_N$ of the
lead to the voltage source $V_N$. For this case of an Ohmic
environment one has\cite{Ingold}
\begin{table}[t!]
\begin{center}
    \begin{tabular}{rcll}
    $\Delta$&=&200$~\mu$eV & superconducting gap\\
    $C_N$&=&1~fF & capacitance of noise generating JJ\\
    $R_t$&=&30~k$\Omega$ & tunneling resistance of noise generating JJ\\
    $R_N$&=&$200~\Omega$ & lead resistance of non-Gaussian noise branch\\
    $R_B$&=&$200~\Omega$ & lead resistance of biasing control branch\\
    \end{tabular}
    \caption[]{\label{tab:Parameters}Circuit parameters used for the
    calculations of the quantities displayed in Figs.\ \ref{fig:IVplot} and \ref{fig:effective}}
\end{center}
\end{table}
\begin{figure}
    \begin{center}
    \includegraphics[width=\columnwidth,draft=false]{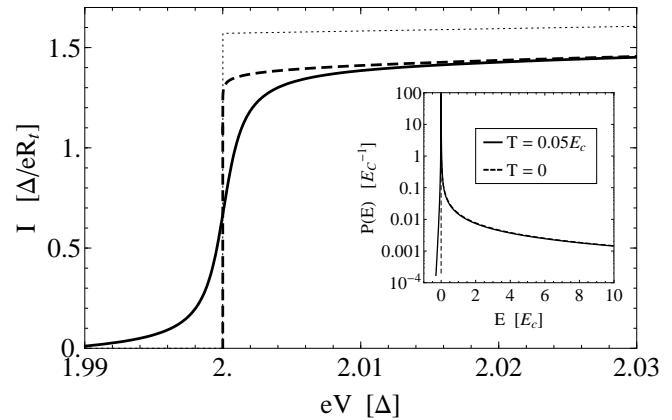}
    \caption{\label{fig:IVplot} $I$-$V$-characteristic of the noise generating JJ at
        $T$=50~mK (solid line) and $T=0$~K (dashed line) in presence of DCB. The dotted line shows the
        step-like discontinuity for a JJ without DCB.
        The inset displays the $P(E)$ function.}
\end{center}
\end{figure}
\begin{eqnarray}
\label{eq:PofE}
  P(E)&=&\int_{-\infty}^{\infty} \frac{dt}{2\pi\hbar}\, e^{iEt/\hbar}\\
  && \times \exp\left[2\int_{-\infty}^{\infty}
  \frac{d\omega}{\omega}\frac{\mbox{Re}Z_{\rm eff}(\omega)}{R_K}
  \frac{e^{-i\omega t}-1}{1-e^{-\hbar\omega/k_{\rm B}T}}
  \right],\nonumber
\end{eqnarray}
where $Z_{\rm eff}(\omega)=R_N/(1+i\omega R_NC_N)$ is the
effective impedance and $R_K=h/e^2$. For the set of parameters
given in Tab.\ \ref{tab:Parameters}, the resulting
$I$-$V$-characteristics of the noise generating JJ is shown in
Fig.\ \ref{fig:IVplot}. The inset shows the function $P(E)$.

The cumulants of the quasiparticle tunneling current are readily
determined from the tunneling rates (\ref{eq:tunnelrate}). One has
\cite{LevitovReznikov}
\begin{eqnarray}
    \CC_{j}^{\rm qp} &=& e^j\left[\Gamma_{\rightarrow}+(-1)^j\Gamma_{\leftarrow}\right]
\end{eqnarray}
which implies, in particular, $\CC_3^{\rm qp}=e^2\CC_1^{\rm qp}$.
Although $R_N$ has a significant effect on the noise cumulants via
DCB, in the region $eV_N>2\Delta$, its influence on the voltage
across the noise generating JJ can safely be neglected since
$R_N/R_t\ll 1$. Hence, the modifications of this voltage are
essentially due to the detector feedback of order $V_J/V_N$
discussed above. Accordingly, in Eqs.\ (\ref{eq:Teff}) and
(\ref{eq:C3eff}), we can put $\CC_j^{N}\simeq\CC_{j}^{\rm
qp}(V_N)$.

\begin{figure}
    \begin{center}
        \includegraphics[width=\columnwidth,draft=false]{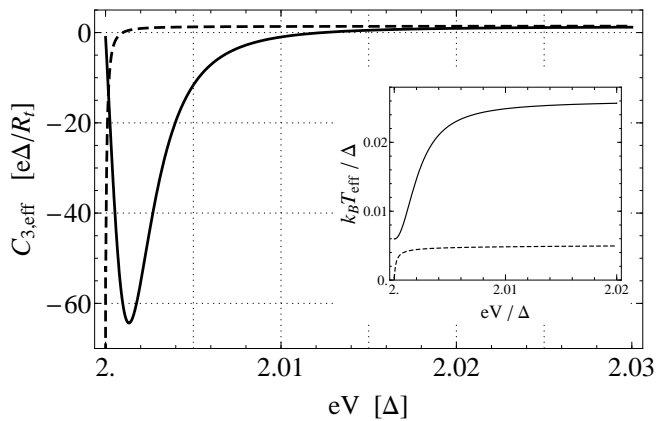}
      \caption{\label{fig:effective} The effective third cumulant for $T$=50~mK (solid line) and $T=0$~K (dashed line).
      The inset shows the effective temperature.}
  \end{center}
\end{figure}

The highly nonlinear behavior of the cumulants for $eV\sim
2\Delta$ strongly affects the effective temperature $\Teff$ and
the effective third cumulant $\Ceff$. Both quantities are shown in
Fig.\ \ref{fig:effective} as functions of the applied voltage,
again for the representative set of parameters given in Tab.\
\ref{tab:Parameters}. While the skewness of the noise $\CC_3$ is
always positive for positive $I_t$, the rate asymmetry
(\ref{asy}), which is proportional to $\Ceff$, passes through zero
as $V_N$ is decreased and it takes large negative values for
voltages slightly above $2\Delta/e$. This change of sign
constitutes a pronounced feedback effect that should be easily
observable experimentally.

In summary, we have shown that feedback of the JJ noise detector
on the noise generating device can be very pronounced if the
device under investigation is highly nonlinear. In particular,
another JJ as noise source should allow for a detailed
experimental study of this effect, which leads to a strongly
enhanced asymmetry of the switching rate of the detector JJ and
even a change of sign of the apparent noise skewness $\Ceff$ as
the voltage applied to the noise generating junction approaches
twice the superconducting gap. The size of the feedback
corrections sensitively depends on DCB of quasiparticle
tunnelling.

The authors wish to thank F.S. Bergeret, D. Esteve, Q. Le Masne,
A. Levy-Yeyati, H. Pothier and C. Urbina for helpful discussions.
Financial support was provided by the nanoscience program of the
European Research Area (ERA).

\end{document}